\documentclass[12pt,a4paper,oneside]{amsart}

\usepackage{amsmath}
\usepackage{latexsym}
\usepackage{amsfonts}
\usepackage{amssymb}
\usepackage{color}
\usepackage{bbm,dsfont}
\usepackage{graphicx}

\newcommand{\R}{\mathbb R} 
\newcommand{\C}{\mathbb C} 
\newcommand{\half}{\tfrac{1}{2}} 
\renewcommand{\mod}[1]{|#1|} 

\newcommand{\ip}[2]{\left\langle\,#1\,|\,#2\,\right\rangle} 
\newcommand{\no}[1]{\left\|#1\right\|} 
\newcommand{\tr}[1]{\textrm{tr}\left[ #1 \right]} 

\newcommand{\va}{\mathbf{a}} 
\newcommand{\vb}{\mathbf{b}} 
\newcommand{\vp}{\mathbf{p}} 
\newcommand{\vq}{\mathbf{q}} 
\newcommand{\vr}{\mathbf{r}} 
\newcommand{\vsigma}{\boldsymbol{\sigma}} 

\newcommand{\A}{\mathcal{A}}
\newcommand{\B}{\mathcal{B}}
\renewcommand{\P}{\mathcal{P}}
\newcommand{\Q}{\mathcal{Q}}

\begin{document}

\title{APPROXIMATE JOINT MEASURABILITY OF SPIN ALONG TWO DIRECTIONS}

\author{Teiko Heinosaari, Peter Stano, and Daniel Reitzner}
\address{Research Center for Quantum Information, Slovak Academy of Sciences, D\'ubravsk\'a cesta 9, 845 11 Bratislava, Slovakia}

\email{temihe@utu.fi; peter.stano@savba.sk; daniel.reitzner@savba.sk}

\begin{abstract}
We study the existence of jointly measurable POVM approximations to two non-commuting sharp spin observables. We compare two different ways to specify optimal approximations.
\end{abstract}

\maketitle
\renewcommand{\figurename}{\footnotesize {Figure}}

\section{Introduction}

Joint measurability for sharp observables is equivalent to the commutativity of the corresponding selfadjoint operators. The question that we study here is the following: having two non-commuting sharp spin observables (which are thus not jointly measurable), what is the closest approximation, in the form of two positive operator valued measures (POVMs), such that these two POVMs are jointly measurable. We show when such approximations exist as regions of allowed points in a suitably chosen space of parameters. This approach allows us, for example, to quantify how far we have to go from the original sharp observables to get jointly measurable approximations. Futhermore, optimal approximations can be identified as those laying on the boundary of such regions.

\section{Statement of the problem}

Let $\P$ and $\Q$ be two observables corresponding to sharp measurements of spin in the directions $\vp$ and $\vq$, respectively. They are described by selfadjoint operators \hbox{$\sigma_\vp\equiv\vp\cdot\vsigma$} and \hbox{$\sigma_\vq\equiv\vq\cdot\vsigma$}, with $\no{\vp}=\no{\vq}=1$. Alternatively, and for our purposes more conveniently, these observables can be described by two outcome PVMs (projection valued measures). Then $\P$ is described by a mapping \hbox{$1\mapsto P$}, \hbox{$-1\mapsto I-P$} with $P$ being the projection \hbox{$P=\half (I+\vp\cdot\vsigma)$}, and $\Q$ is similarly described by a PVM corresponding to projections \hbox{$Q=\half (I+\vq\cdot\vsigma)$} and $I-Q$. We denote by $\theta$ the angle between $\vp$ and $\vq$ and we assume that $0<\theta\leq 90^\circ$.

A joint measurement for $\P$ and $\Q$ is defined as a measurement with four outcomes corresponding to four possible pairs of $\P$ and $\Q$ outcomes, $(\pm 1,\pm 1)$. In addition, it is required that the measurement outcome statistics for $\P$ ($\Q$) measured alone can be obtained from the joint measurement by disregarding (summing through all possible) outcomes for $\Q$ ($\P$). Thus, $\P$ and $\Q$ are jointly measurable if there exist four operators $G_{++}, G_{+-}, G_{-+}, G_{--}$ such that 
\begin{equation}\label{joint measurability}
P=G_{++} + G_{+-}\, ,\quad Q=G_{++} + G_{-+} \, .
\end{equation}
These operators must form a POVM, hence they are positive and satisfy \hbox{$\sum_{ij} G_{ij}=I$}. It is well known that $\P$ and $\Q$ are jointly measurable if and only if $P$ and $Q$ commute, which is the case when \hbox{$\vp=\pm\vq$}. Thus, if \hbox{$\vp\neq\pm\vq$}, a joint measurement can only approximate $\P$ and $\Q$.

We are looking for observables $\A$ and $\B$ such that they are jointly measurable and  can be taken as approximations to $\P$ and $\Q$. An observable $\A$ with two outcomes $\pm 1$ is described by a POVM $1\mapsto A$, $-1\mapsto I-A$, where $A$ is an operator on $\C^2$ satisfying $O\leq A \leq I$. It can be parametrized by four real parameters,
\begin{equation}\label{a}
A=\frac{1}{2} (\alpha I+\va\cdot\vsigma),\quad \no{\va}\leq\alpha\leq 2-\no{\va} \, .
\end{equation} 
If $\alpha=1$, then the condition in (\ref{a}) reduces to \hbox{$\no{\va}\leq 1$}.
In a similar way, $\B$ is described by a POVM corresponding to operator \hbox{$B=\half (\beta I + \vb\cdot\vsigma)$} with \hbox{$\no{\vb}\leq\beta\leq 2- \no{\vb}$}. 

For sharp observables joint measurability is equivalent to commutativity. The decision whether two observables are jointly measurable is, however, more involved for observables in general. A general characterization of joint measurability of $\A$ and $\B$ is, up to authors knowledge, an open problem. However, for the task under consideration the following result \cite{Busch86} of Busch will be enough: if $\alpha=\beta=1$, then the necessary and sufficient condition for $\A$ and $\B$ to be jointly measurable is
\begin{equation}\label{coex}
\no{\va-\vb}+\no{\va+\vb}\leq 2 \, .
\end{equation}
If $\alpha$ and $\beta$ are not equal to 1, this condition is still necessary for joint measurability \cite{BuHe07}.

\begin{figure}
\centerline{\includegraphics[scale=0.7]{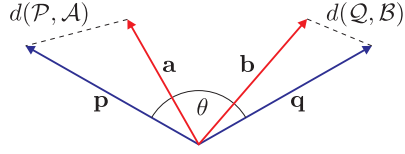}}
\caption{\label{fig:definition}The observables $\P$ and $\Q$ are parametrized by unit vectors $\vp$ and $\vq$ making an angle $\theta$. Two jointly measurable POVM approximations $\A$ and $\B$ are parametrized by vectors $\va$ and $\vb$. The smaller is the distance $d(\P,\A)$, the better $\A$ approximates $\P$.}
\end{figure}

We will quantify how well $\A$ approximates $\P$ using two different distances $d(\P,\A)$ in the next sections. We then solve the following problem: for a fixed distance\footnote{Observable $\A$ is not fixed - we allow all $\A$ which have the fixed distance from $\P$.} $d(\P,\A)$, find the smallest possible distance $d(\Q,\B)$ such that $\A$ and $\B$ are jointly measurable. The task is summarized in Fig.~\ref{fig:definition}.

\section{Statistical distance}\label{sec:statistical}

One possible way to quantify the distance between $\A$ and $\P$ is to compare the probabilities that these observables give for states. There are at least two reasonable choices: we can calculate the worst possible deviation or the average deviation between the probability distributions on the outcomes. For the worst possible deviation we get
\begin{equation}\label{worst}
\textrm{sup}_{\rho} \mid \tr{\rho P}  - \tr{\rho A} \mid =\half\no{\vp-\va}+\half\mod{1-\alpha}\, .
\end{equation}
On the other hand, the average deviation can be calculated using pure states $\psi_\vr$ parametrized by the points $\vr$ in the unit sphere in $\R^3$, and we get\footnote{For $\vp=\va$ the right hand side of (\ref{average}) is $\half\mod{1-\alpha}$ and the following analysis still holds.}
\begin{equation}\label{average}
\frac{1}{4\pi} \int_{\vr\in S^2} \mid \ip{\psi_\vr}{P\psi_\vr} - \ip{\psi_\vr}{A\psi_\vr} \mid d\vr=\frac{1}{4}\no{\vp-\va}+\frac{(1-\alpha)^2}{4\no{\vp-\va}}\, .
\end{equation}

Clearly, in both cases the choice $\alpha=1$ makes the distance smallest independently on $\va$. Moreover, as shown in Ref.~\cite{BuHe07} the joint measurability condition for $\A$ and $\B$ is the least restrictive when $\alpha=\beta=1$. Therefore, we can restrict ourselves to this case in the search for optimal approximations and in the rest of this section we set $\alpha=\beta=1$ .

With $\alpha=1$ the two different ways (\ref{worst}) and (\ref{average}) to compare $A$ to $P$ give the same value, up to a factor $\half$. As the scale of the distance is not important for our purposes, we define
\begin{equation} \label{distance}
d_s(\P,\A):=\frac{1}{2}\no{\vp-\va} \, .
\end{equation}

\begin{figure}
\centerline{\includegraphics[scale=0.7]{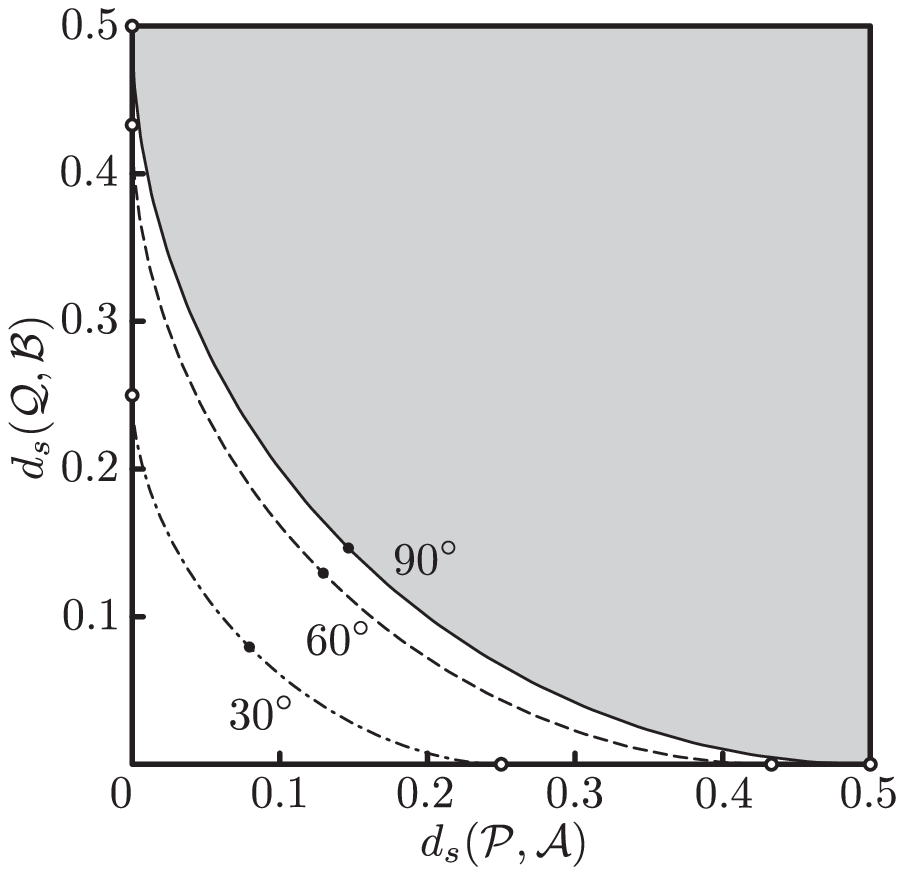}}
\caption{\label{fig:solution} Curves represent boundaries for jointly measurable approximations $\A$ and $\B$ of $\P$ and $\Q$ for different choices of angle $\theta$ between the vectors $\vp$ and $\vq$. The gray-shaded region represents the area where there exist jointly measurable pairs $\A$ and $\B$ for $\theta=90^\circ$.}
\end{figure}

With the definition of distance in Eq.~\eqref{distance}, the possible jointly measurable approximations to sharp observables $\P$, $\Q$ are shown in Fig.~\ref{fig:solution}. Taking the case of $\theta=90^\circ$ first, the shaded area represents points where jointly measurable approximations $\A$ and $\B$ exist, with the distance from $\P$ and $\Q$ given on the $x$ and $y$ axis, respectively. The non-shaded area in the left down corner represents points where the requirement on the approximation is too high such that no jointly measurable approximations $\A$ and $\B$ exist. Moving from the left down corner to the right we relax our requirements on how closely $\A$ approximates $\P$ until finaly we reach the boundary of the shaded region where some jointly measurable approximations exist. Similarly, moving up in the figure, we allow less strict approximation for $\Q$. The boundary of the shaded area thus represents the optimal choice of jointly measurable approximations. Moving along the boundary curve one sees an expected trade off -- if we require $\A$ to better approximate $\P$, we have to loosen our requirements on $\B$ to preserve joint measurability. We also give boundary curves for the cases $\theta=60^\circ$ and $30^\circ$. All these curves have been calculated numerically. Again, the region up right from these boundaries is the area where jointly measurable approximations exist. With decreasing $\theta$, the boundary moves towards the left down corner, which we could interpret as the two sharp observables $\P$ and $\Q$ becoming more easily jointly measurable. In the limit case $\theta=0^\circ$, the whole area would be shaded -- we can find jointly measurable approximations with any desirable accuracy since the two sharp observables $\P$ and $\Q$ are then the same.

The same problem has been studied analytically in Ref.~\cite{BuHe07} where the optimal solution was found in the case $d_s(\P,\A)=d_s(\Q,\B)$; the corresponding points in Fig.~\ref{fig:solution} are denoted by black dots. It was proved that the optimal $\A$ and $\B$ are then given by 
\begin{equation}
\va=\lambda \frac{\vp+\vq}{\no{\vp+\vq}}+ (1-\lambda) \frac{\vp-\vq}{\no{\vp-\vq}},\quad \vb=\lambda \frac{\vp+\vq}{\no{\vp+\vq}}- (1-\lambda) \frac{\vp-\vq}{\no{\vp-\vq}}\, ,
\end{equation}
where $\lambda=\half (1+\cos \frac{\theta}{2} - \sin \frac{\theta}{2} )$. If $\theta=90^\circ$, then $\va=\frac{1}{\sqrt{2}}\vp$ and $\vb=\frac{1}{\sqrt{2}}\vq$. Otherwise $\va$ and $\vb$ are not parallel to $\vp$ and $\vq$ but they are somewhere between them as sketched in Fig.\ref{fig:definition}.

\section{Root-mean-square noise}\label{sec:rms}

In a series of recent articles \cite{Ozawa04a,Ozawa04b,Ozawa05} Ozawa has investigated noise and disturbance in quantum measurements. Assume that we try to measure an observable $\P$ (described by a selfadjoint operator $\sigma_\vp$) but we actually perform a measurement of $\A$. A precise measurement of $\P$ would mean that $\A=\P$. Otherwise it is thought that we have a measurement of $\P$ with some noise. Ozawa defines the so-called root-mean-square noise $\epsilon_{rms}$ by the following formula \cite{Ozawa04a}:
\begin{equation}
\epsilon_{rms}(\P,\A;\rho):=\tr{\left( \A[2]-\A[1]^2 +(\sigma_\vp-\A[1])^2 \right) \rho}^{\half} \, .
\end{equation}
Here $\rho$ is an input state and $\A[1]$ and $\A[2]$ are the first and the
second moment operators of $\A$, respectively.

Let now $\A$ be a two outcome observable with outcomes $\pm1$ and $A$ as in (\ref{a}). A short calculation shows that if $\alpha=1$ then $\epsilon_{rms}(\P,\A;\rho)$ is a state independent number and we can define
\begin{equation}\label{drms}
d_{rms}(\P,\A):=\epsilon_{rms}(\P,\A;\rho)=\sqrt{ 2(1-\va\cdot\vp )}\, .
\end{equation}
Generally (i.e. $\alpha\neq 1$) we define $d_{rms}$ to be the worst deviation over all input states. However, it can be shown that the optimal solutions are to be found among the operators having $\alpha=1$, so we again restrict our study to this case.

\begin{figure}
\centerline{\includegraphics[scale=0.7]{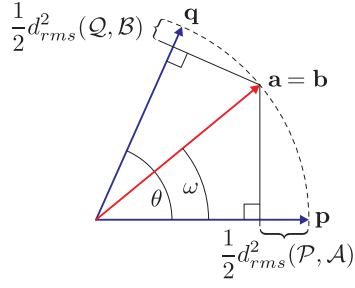}}
\caption{\label{fig:ozawa} 
For a fixed distance $d_{rms}(\P,\A)$, the smallest distance $d_{rms}(\Q,\B)$ is obtained when $\va=\vb$ is the unit vector shown in the picture.}
\end{figure}

\begin{figure}[t!]
\centerline{\includegraphics[scale=0.7]{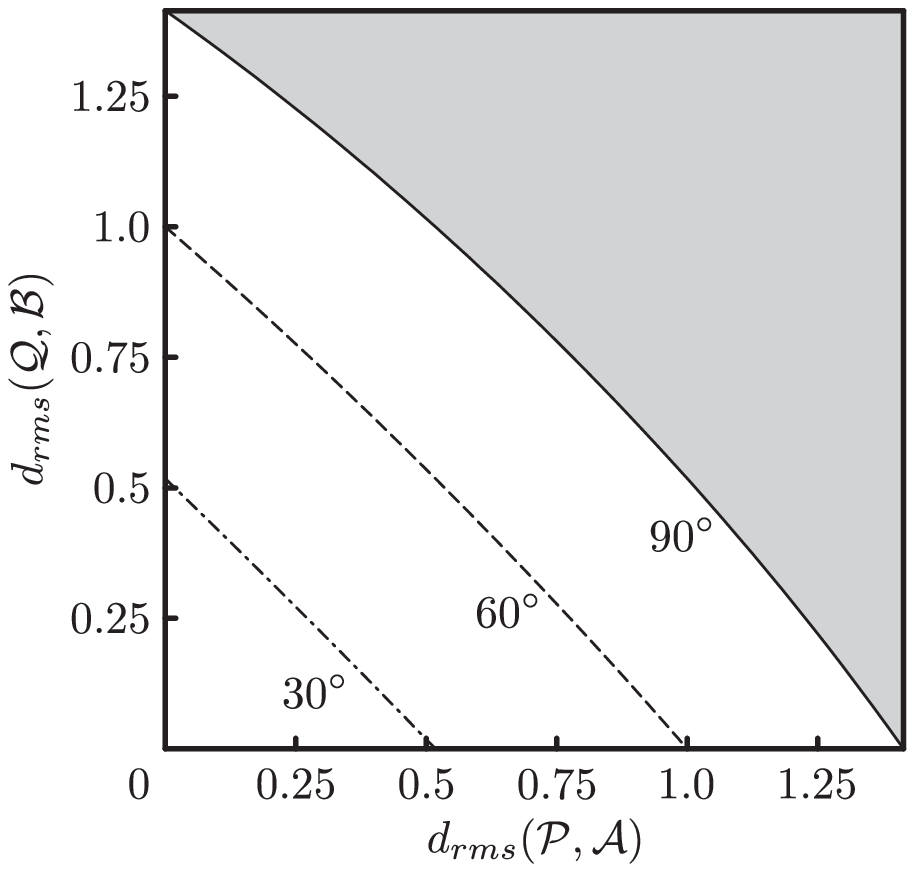}}
\caption{\label{fig:ozawa2}  Curves represent boundaries for jointly measurable approximations $\A$ and $\B$ of $\P$ and $\Q$ for different choices of angle $\theta$ between the vectors $\vp$ and $\vq$. The gray-shaded region represents the area where there exist jointly measurable pairs $\A$ and $\B$ for $\theta=90^\circ$.
}
\end{figure}

Assume that $d_{rms}(\P,\A)$ is fixed and we are looking for the smallest possible number $d_{rms}(\Q,\B)$ such that $\A$ and $\B$ are jointly measurable. According to Eq. (\ref{drms}), vectors $\va$ ($\vb$) ending on the line perpendicular to $\vp$ ($\vq$) have the same distance $d_{rms}(\P,\A)$ ($d_{rms}(\Q,\B)$). Going through all the possible pairs $\va$ and $\vb$ satisfying the joint measurability condition (\ref{coex}), it can be shown that the optimal situation corresponds to the choice $\va=\vb$, where $\va$ is a unit vector and the angle $\omega$ between $\va$ and $\vp$ is $\omega=\arccos(1-\half d_{rms}(\P,\A)^2)$; see Fig.~\ref{fig:ozawa}. This solution means that in order to perform an optimal joint measurement, we measure spin in the direction $\va=\vb$. This is then regarded as an approximate version of both $\sigma_\vp$ and $\sigma_\vq$. Combinations of $d_{rms}(\P,\A)$ and $d_{rms}(\Q,\B)$, for which jointly measurable approximations $\A$ and $\B$ exist, are shown in Fig.~\ref{fig:ozawa2}.

The difference between the conclusions obtained here and in Section \ref{sec:statistical} can be explained by observing that while $d_{s}$ quantifies the accuracy of how $\A$ approximates $\P$,  $d_{rms}$ contains also a term which depends only on $\A$. Indeed,
\begin{equation}
d_{rms}(\P,\A)^2=\no{\vp-\va}^2 + 1-\no{\va}^2\, .
\end{equation}
The first part $\no{\vp-\va}^2$ is related to the statistical distance $d_s$ while the second part $1-\no{\va}^2$ can be interpreted as a quantification of the intrinsic unsharpness of $\A$. 

\section*{Acknowledgements}

This work was supported by projects CONQUEST, QAP and APVV.

\vfill
\hrule
\end{document}